\documentclass[%
 reprint,
 amsmath,amssymb,
 aps,
 prl,
floatfix,
]{revtex4-2}

\usepackage{romanbar}
\usepackage{xcolor}
\usepackage{graphicx}
\usepackage{dcolumn}
\usepackage{bm}
\usepackage{SIunits}
\usepackage{array}
\usepackage{subfigure}
\usepackage{multirow}%

\usepackage{hyperref} 

\usepackage{url}

\begin{document}


\title{COSINE-100 Full Dataset Challenges the Annual Modulation Signal of DAMA/LIBRA}

\author{Nelson~Carlin$^{1}$}
\author{Jae~Young~Cho$^{2, 3}$}
\author{Jae~Jin~Choi$^{4, 2}$}
\author{Seonho~Choi$^{4}$}
\author{Anthony~C.~Ezeribe$^{5}$}
\author{Luis~Eduardo~Fran{\c c}a$^{1}$}
\author{Chang~Hyon~Ha$^{6}$}
\author{In~Sik~Hahn$^{7, 8, 9}$}
\author{Sophia~J.~Hollick$^{10}$}
\author{Eunju~Jeon$^{2, 9}$}
\author{Han~Wool~Joo$^{4}$}
\author{Woon~Gu~Kang$^{2}$}
\author{Matthew~Kauer$^{11}$}
\author{Bongho~Kim$^{2}$}
\author{Hongjoo~Kim$^{3}$}
\author{Jinyoung~Kim$^{6}$}
\author{Kyungwon~Kim$^{2}$}
\author{SungHyun~Kim$^{2}$}
\author{Sun~Kee~Kim$^{4}$}
\author{Won~Kyung~Kim$^{9, 2}$}
\author{Yeongduk~Kim$^{2, 9}$}
\author{Yong-Hamb~Kim$^{2, 9}$}
\author{Young~Ju~Ko$^{12}$} \email{yjkophys@jejunu.ac.kr}
\author{Doohyeok~Lee$^{3}$}
\author{Eun~Kyung~Lee$^{2}$}
\author{Hyunseok~Lee$^{9, 2}$}
\author{Hyun~Su~Lee$^{2, 9}$} \email{hyunsulee@ibs.re.kr}
\author{Hye~Young~Lee$^{7}$}
\author{In~Soo~Lee$^{2}$}
\author{Jaison~Lee$^{2}$}
\author{Jooyoung~Lee$^{3}$}
\author{Moo~Hyun~Lee$^{2, 9}$}
\author{Seo~Hyun~Lee$^{9, 2}$}
\author{Seung~Mok~Lee$^{4}$} \email{physmlee@gmail.com}
\altaffiliation[Currently affiliated with ]{Physics Department, Carnegie Mellon University}
\author{Yujin~Lee$^{6}$}
\author{Douglas~S.~Leonard$^{2}$}
\author{Nguyen~Thanh~Luan$^{3}$}
\author{Vitor~Hugo~de~Almeida~Machado$^{1}$}
\author{Bruno~B.~Manzato$^{1}$}
\author{Reina~H.~Maruyama$^{10}$}
\author{Robert~J.~Neal$^{5}$}
\author{Stephen~L.~Olsen$^{2}$}
\author{Byung~Ju~Park$^{9, 2}$}
\author{Hyang~Kyu~Park$^{13}$}
\author{Hyeonseo~Park$^{14}$}
\author{Jong-Chul~Park$^{15}$}
\author{Kangsoon~Park$^{2}$}
\author{Se~Dong~Park$^{3}$}
\author{Ricardo~L.~C.~Pitta$^{1}$}
\author{Hafizh~Prihtiadi$^{16}$}
\author{Sejin~Ra$^{2}$}
\author{Carsten~Rott$^{17, 18}$}
\author{Keon~Ah~Shin$^{2}$}
\author{David~F.~F.~S.~Cavalcante$^{1}$}
\author{Min~Ki~Son$^{15}$}
\author{Neil~J.~C.~Spooner$^{5}$}
\author{Lam~Tan~Truc$^{3}$}
\author{Liang~Yang$^{19}$}
\author{Gyun~Ho~Yu$^{17, 2}$}

\affiliation{\small$^{1}$Physics Institute, University of S\~{a}o Paulo, 05508-090, S\~{a}o Paulo, Brazil}
\affiliation{\small$^{2}$Center for Underground Physics, Institute for Basic Science (IBS), Daejeon 34126, Republic of Korea}
\affiliation{\small$^{3}$Department of Physics, Kyungpook National University, Daegu 41566, Republic of Korea}
\affiliation{\small$^{4}$Department of Physics and Astronomy, Seoul National University, Seoul 08826, Republic of Korea}
\affiliation{\small$^{5}$Department of Physics and Astronomy, University of Sheffield, Sheffield S3 7RH, United Kingdom}
\affiliation{\small$^{6}$Department of Physics, Chung-Ang University, Seoul 06973, Republic of Korea}
\affiliation{\small$^{7}$Center for Exotic Nuclear Studies, Institute for Basic Science (IBS), Daejeon 34126, Republic of Korea}
\affiliation{\small$^{8}$Department of Science Education, Ewha Womans University, Seoul 03760, Republic of Korea}
\affiliation{\small$^{9}$IBS School, University of Science and Technology (UST), Daejeon 34113, Republic of Korea}
\affiliation{\small$^{10}$Department of Physics and Wright Laboratory, Yale University, New Haven, CT 06520, USA}
\affiliation{\small$^{11}$Department of Physics and Wisconsin IceCube Particle Astrophysics Center, University of Wisconsin-Madison, Madison, WI 53706, USA}
\affiliation{\small$^{12}$Department of Physics, Jeju National University, Jeju 63243, Republic of Korea}
\affiliation{\small$^{13}$Department of Accelerator Science, Korea University, Sejong 30019, Republic of Korea}
\affiliation{\small$^{14}$Korea Research Institute of Standards and Science, Daejeon 34113, Republic of Korea}
\affiliation{\small$^{15}$Department of Physics and IQS, Chungnam National University, Daejeon 34134, Republic of Korea}
\affiliation{\small$^{16}$Department of Physics, Universitas Negeri Malang, Malang 65145, Indonesia}
\affiliation{\small$^{17}$Department of Physics, Sungkyunkwan University, Suwon 16419, Republic of Korea}
\affiliation{\small$^{18}$Department of Physics and Astronomy, University of Utah, Salt Lake City, UT 84112, USA}
\affiliation{\small$^{19}$Department of Physics, University of California San Diego, La Jolla, CA 92093, US}

\collaboration{COSINE-100 Collaboration} \noaffiliation

\begin{abstract}
For over 25 years, the DAMA/LIBRA collaboration has claimed to observe an annual modulation signal, suggesting the existence of dark matter interactions.
However, no experiment employing different target materials has observed a dark matter signal consistent with their result.
To address this puzzle, the COSINE-100 collaboration conducted a model-independent test using sodium iodide crystal detectors, the same target material as DAMA/LIBRA. Analyzing data collected over 6.4 years by the effective mass of 61.3\,kg, with improved energy calibration and time-dependent background modeling, we found no evidence of an annual modulation signal, challenging the DAMA/LIBRA result with a confidence level greater than 3$\sigma$. This finding represents a substantial step toward resolving the long-standing debate surrounding DAMA/LIBRA’s dark matter claim, indicating that the observed modulation is unlikely to be caused by dark matter interactions.
\end{abstract}

\maketitle


\section{Introduction}\label{sec1}

Cosmological observations indicate~\ref{sec1} the existence of dark matter, which is thought to constitute the majority of matter in the Universe~\cite{Planck:2018vyg, Bertone:2016nfn}.
A favored explanation for this dark matter is a population of weakly interacting massive particles~(WIMPs)~\cite{PhysRevLett.39.165} that are potentially detectable through their rare interactions with atomic nuclei in terrestrial experiments~\cite{Goodman:1984dc}. 
Over the past three decades, extensive efforts have been devoted to detecting WIMPs directly by observing their elastic scattering off atomic nuclei.
However, no experiment has yet produced conclusive evidence for their existence~\cite{Billard:2021uyg, ParticleDataGroup:2022pth}. 

One of the most intriguing claims for dark matter detection comes from the DAMA/LIBRA experiment, which has reported a periodic modulation in its event rate for over 25 years~\cite{Bernabei:1998fta, Bernabei:2013xsa, Bernabei:2018yyw, Bernabei:2021kdo}.
According to theoretical predictions concerning WIMP dark matter within the standard halo model~\cite{Savage:2008er,COSINE-100:2019brm}, event rates in a direct detection experiment are expected to exhibit an annual modulation due to the relative motion of the Earth with respect to the galactic halo~\cite{Lewin:1995rx,freese1987,Freese:2012xd,Froborg:2020tdh}.
Therefore, DAMA/LIBRA’s observation of a modulation signal has been widely interpreted as potential evidence of dark matter interactions. However, this claim remains unverified by other experiments.
A key challenge in evaluating the DAMA/LIBRA signal is determining whether the observed modulation is genuinely caused by dark matter or by experimental or systematic effects.
Resolving such discrepancies requires model-independent experimental comparisons, where different groups attempt to reproduce the result using the same target material, independent experimental setups, and distinct analysis frameworks.
The ideal approach to such comparisons is to minimize differences in detector response and systematic uncertainties while ensuring that the methodologies used are robust, transparent, and unbiased.
These principles provide the foundation for rigorously testing DAMA/LIBRA’s claim.

DAMA/LIBRA observed a modulation amplitude of 0.0191$\pm$0.0020\,counts/day/kg/keV in the 1--3\,keV low-energy range, with high statistical significance levels in several energy ranges above 10$\sigma$~\cite{Bernabei:1998fta, Bernabei:2013xsa, Bernabei:2018yyw, Bernabei:2021kdo}.
If this signal is attributed to WIMP interactions, it contradicts findings from other direct detection experiments using different target materials\cite{SuperCDMS:2017mbc,cresst730kg,DarkSide:2018kuk,PICO:2019vsc,PandaX-4T:2021bab,XENONCollaboration:2023orw,LZ:2022ufs}, which have found no evidence of dark matter within the parameter space allowed by DAMA/LIBRA.
This discrepancy is especially pronounced in annual modulation analyses with liquid xenon detectors, which show no modulation signals~\cite{XENON:2017nik,LUX:2018xvj,XMASS:2018koa}.
However, none of these other experiments have utilized sodium iodide crystals as DAMA/LIBRA, leaving the question of whether the target material plays a critical role in detecting potential WIMP signals.

To address this question, sodium iodide-based experiments such as COSINE-100~\cite{Adhikari:2017esn} and ANAIS-112~\cite{Amare:2019jul} have been launched to independently verify DAMA/LIBRA's findings using similar detectors~\cite{kwkim15, PhysRevD.110.043010, Fushimi:2021mez, Calaprice:2021yml}. 
COSINE-100 has conducted model-dependent analyses using low-energy spectra to search for WIMP signals, finding no evidence consistent with DAMA/LIBRA when interpreted within the standard halo model~\cite{Adhikari:2018ljm,COSINE-100:2021xqn}. 
The ANAIS-112 experiment has published model-independent results from three years of data, reporting a modulation amplitude of $-0.0034 \pm 0.0042$\,counts/day/kg/keV in the 1--6\,keV energy range, which is consistent with no modulation and incompatible with the DAMA/LIBRA's findings~\cite{Amare:2021yyu}.
Recent improvements in ANAIS-112's analysis have further enhanced their sensitivity to 2.8$\sigma$ with respect to the DAMA/LIBRA signal in both the 1--6\,keV and 2--6\,keV ranges~\cite{Coarasa2024}.
Similarly, COSINE-100's three-year data analysis reported no modulation~\cite{COSINE-100:2021zqh}, though a slightly positive mean value introduces some uncertainty.
The previous results of the two experiments are summarized in Tables~\ref{tab:PhaseFixedFittingResults} and \ref{tab:PhaseFixedFittingResultsNr}.
While these findings are promising, neither experiment has yet achieved the statistical sensitivity required to definitively confirm or refute DAMA/LIBRA's results, highlighting the need for higher statistics.

One challenge in comparing these experiments lies in the nuclear recoil quenching factors (QFs), which represent the scintillation light yield from sodium or iodine recoils relative to electron recoils at the same energy.
DAMA/LIBRA has reported higher QFs for their NaI(Tl) crystals~\cite{Bernabei:2020mon} compared to the measurement from COSINE-100~\cite{Lee:2024unz}.
Although recent measurements from other groups, including ANAIS-112 and SABRE, are consistent with each other and with COSINE-100's findings~\cite{Collar:2013gu, Xu:2015wha, Cintas:2024pdu, Lee:2024unz}, the possibility that DAMA/LIBRA's crystals exhibit higher QFs remains unresolved.
This discrepancy suggests that previous model-independent comparisons~\cite{Amare:2021yyu, COSINE-100:2021zqh} between experiments may not fully account for dark matter-nuclei interactions~\cite{Bernabei:2020mon}.
ANAIS-112 recently examined this issue, reporting no modulation and finding a 2.3$\sigma$ inconsistency with DAMA/LIBRA's signal when accounting for QFs~\cite{Coarasa2024}. 

Here, we present results from a search for dark matter-induced annual modulation in the full COSINE-100 dataset, corresponding to 5.8 years of high quality data with a reduced energy threshold of 0.7\,keV from 1\,keV~\cite{COSINE-100:2024SEL}.
Unlike our previous analyses~\cite{COSINE-100:2019lgn, COSINE-100:2021zqh}, we adopt an energy calibration as closely aligned with DAMA/LIBRA's calibration as possible to ensure a rigorous comparison. 
We also consider a range of QF scenarios to address the potential effects of differing nuclear recoil responses. 
This study aims to provide a comprehensive and statistically robust test of the DAMA/LIBRA claim within the framework of NaI(Tl)-based dark matter searches. 


\section{Experimental setup}
The COSINE-100 detector comprises a 106\,kg array of eight low-background thallium-doped sodium iodide crystals, each optically coupled to two photomultiplier tubes~(PMTs).
These sodium iodide crystal assemblies are submerged in 2,200 liters of liquid scintillator, enabling the identification and subsequent reduction of radioactive backgrounds observed by the crystals~\cite{Adhikari:2020asl}.
The liquid scintillator is surrounded by copper, lead, and plastic scintillator to reduce background contributions from external radiation and cosmic-ray muons~\cite{Prihtiadi:2017inr}.
Further details of the setup are provided elsewhere~\cite{Adhikari:2017esn} (see also Fig.~\ref{fig:cosinedetector}).

COSINE-100 was installed at the Yangyang Underground Laboratory~(Y2L) in Korea, located underground at a water-equivalent depth of approximately 1,800 meters~\cite{Prihtiadi:2017inr, COSINE-100:2020jml}, and conducted physics data-taking operations from October 2016 to March 2023.
Following the completion of data collection at Y2L, the COSINE-100 detector was disassembled and relocated to Yemilab~\cite{Park:2024sio,Kim:2024xyd}, a new, deeper underground laboratory in Korea, and an upgrade of the COSINE-100 is currently being installed at Yemilab~\cite{Lee:2024wzd}. 

\begin{figure*}[t]
    \centering
    \includegraphics[width=1.0\textwidth]{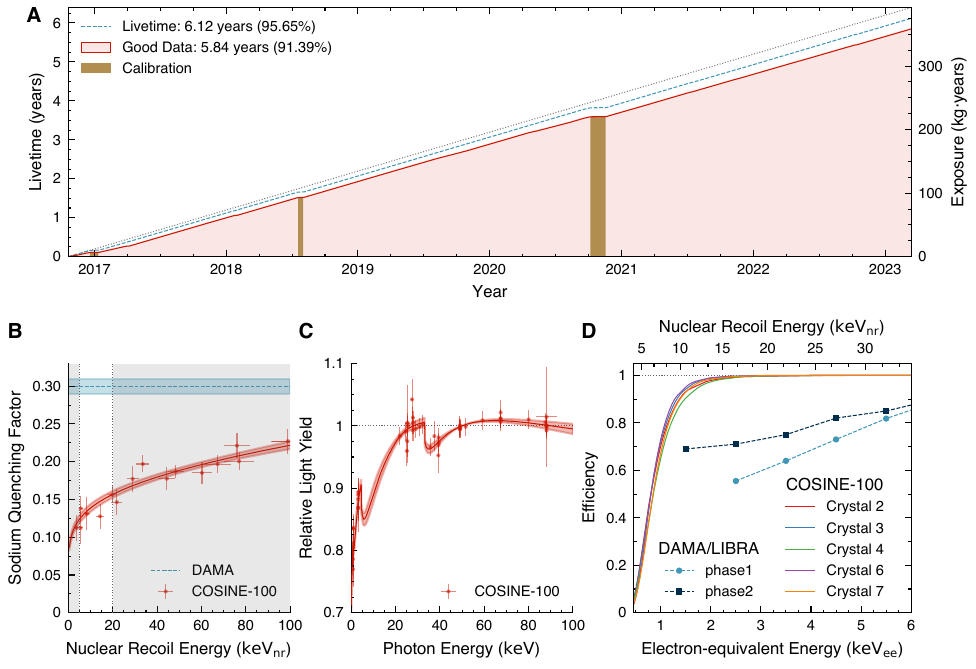}
    \caption{
        {\bf COSINE-100 detector details.}
        ({\bf A}) Operation overview.
        The COSINE-100 detector collected dark matter search data from October 21, 2016 to March 14, 2023, amounting to 6.4 years of operation.
        We analyzed 358.4\,kg$\cdot$years high quality data to search for dark matter-induced modulation signals.
        ({\bf B}) Quenching factor comparison.
        The sodium quenching factors measured by COSINE-100 (red solid line)~\cite{Lee:2024unz} and DAMA (blue dashed line)~\cite{BERNABEI1996757} are displayed, highlighting the signal area from 5\,\ensuremath{{\rm keV_{nr}}} to 20\,\ensuremath{{\rm keV_{nr}}}~(bright area).
        ({\bf C}) Nonproportionality of the COSINE-100 crystals~\cite{COSINE-100:2024log}.
        The light yields for various energy of $\gamma$ and x-rays are displayed relatively to the value at 50\,keV.
        To directly compare with DAMA/LIBRA, \ensuremath{{\rm keV_{ee}}} unit in this article does not incorporate such a nonproportional response.
        ({\bf D}) Event selection efficiencies.
        The event selection efficiencies are compared across the energy range of \ensuremath{{\rm keV_{ee}}}, based on the linear calibration from five COSINE-100 crystals~\cite{COSINE-100:2024SEL}, with those from DAMA/LIBRA-phase1~\cite{Bernabei:2008yh} and DAMA/LIBRA-phase2~\cite{Bernabei:2012zzb}.
    }
    \label{fig:detail}
\end{figure*}

Throughout the 6.4-year data-taking period at Y2L, no notable environmental abnormalities or unstable detector performances were observed, achieving more than 95\% time efficiency for collecting physics data (see Fig.~\ref{fig:detail}A).
For this analysis, we utilize the full dataset of COSINE-100 operations, totaling an effective livetime of 5.8 years.
Eight low-background thallium-doped sodium iodide crystals were operated; however, two of these crystals had low light yields and one had a high PMT-induced noise rate.
These crystals were therefore excluded from this analysis, resulting in a total effective mass of 61.3\,kg~\cite{COSINE-100:2019lgn, COSINE-100:2021xqn, COSINE-100:2021zqh} and an effective exposure of 358\,kg$\cdot$years used for this analysis.

Various monitoring devices were installed in the COSINE-100 detector system to ensure stable data-taking and systematic analysis of the annual modulation~\cite{COSINE-100:2021mlj}.
These devices monitored the temperatures in the tunnel, detector room, and liquid scintillator, as well as humidity, radon levels, high voltages and currents to the PMTs, and thermal and fast neutron rates, along with the data acquisition systems.
Continuous monitoring allowed verification of the stability of the environment and detector in real time.
Additionally, monitoring of internal x-ray peaks at 3.2\,keV and 0.87\,keV from the decay of $^{40}$K and $^{22}$Na, respectively, ensured the stability of low-energy calibration and gain of the sodium iodide detectors (see Fig.~\ref{fig:detail_peak}).

The total scintillation light of an event, the indicator of its energy, was calculated using the integration of the waveform over a 5\,$\mu$s time window.
For a long time, it has been known that the relationship between deposited energy and the scintillation light produced in sodium iodide is nonlinear, which is called the nonproportionality phenomenon.
The nonproportionality in the COSINE-100 crystals was also measured, using internal $\gamma$ and x-ray lines, and external $\gamma$ radiation incident on a sample crystal, as shown in Fig.~\ref{fig:detail}C~\cite{COSINE-100:2024log}.
This enhanced the understanding of the background contributing to the COSINE-100 detector~\cite{COSINE-100:2024Background}.
Previous annual modulation analyses from the COSINE-100 experiment~\cite{COSINE-100:2019lgn, COSINE-100:2021zqh} used a calibrated energy scale that accounted for the nonproportionality of the detectors.
However, the DAMA/LIBRA analysis employed a simple linear energy calibration using various $\gamma$ lines~\cite{Bernabei:2008yh}, resulting in a slightly different energy scale in the low-energy signal regions.
In the analysis presented here, an electron-equivalent energy calibration similar to DAMA/LIBRA~\cite{Bernabei:2008yh} is adopted, with a linear calibration using the 59.5\,keV $\gamma$ line. We employ a unit of apparent energy called kilo-electron-volt electron-equivalent (\ensuremath{{\rm keV_{ee}}}) for DAMA/LIBRA-like linear calibration. 

One notable challenge is the discrepancy between nuclear-recoil QFs of sodium and iodine reported by the DAMA/LIBRA collaboration~\cite{BERNABEI1996757} and recent measurements by other groups~\cite{Collar:2013gu, Xu:2015wha, Cintas:2024pdu, Lee:2024unz}.
The DAMA/LIBRA collaboration measured the response of the sodium iodide crystal to nuclear recoils induced by neutrons from a $^{252}$Cf source.
The measured responses were compared with simulated neutron energy spectra to obtain the constant QF values of 0.3 for sodium and 0.09 for iodine, without energy dependency~\cite{BERNABEI1996757}.

However, the recent measurements by other groups used monochromatic neutron beams with neutron tagging detectors that measure elastically scattered neutrons at a fixed angle relative to the incoming neutron beam direction, providing accurate knowledge of the nuclear recoil energy transferred from incoming neutrons to sodium or iodine nuclei~\cite{Collar:2013gu, Xu:2015wha, Cintas:2024pdu, Lee:2024unz}.
QF values ($\sim$0.13 for sodium and $\sim$0.05 for iodine at 10\,keV of nuclear recoil energy) from these measurements differ substantially from DAMA/LIBRA QF results.
DAMA/LIBRA pointed out that comparisons with other sodium iodide-based experiments did not fully account for their QF results in the case of dark matter-nuclei interactions~\cite{Bernabei:2020mon}.

Here, we consider the possibility of different QF scenarios for the DAMA/LIBRA crystals compared to the COSINE-100 crystals.
Although both sodium and iodine interactions can be considered, we focus on the sodium nuclei due to the strong constraints on the iodine interaction that come from xenon-based dark matter search experiments~\cite{XENONCollaboration:2023orw, LZ:2022ufs}, where xenon has a similar atomic mass number to iodine.
The nuclear-recoil equivalent energy (a unit of \ensuremath{{\rm keV_{nr}}}) considers different QF values between DAMA/LIBRA~\cite{BERNABEI1996757} and COSINE-100~\cite{Lee:2024unz}.
For example, the 2--6\,\ensuremath{{\rm keV_{ee}}} energy region of DAMA/LIBRA corresponds to 6.67--20\,\ensuremath{{\rm keV_{nr}}}, which corresponds to 0.85--3.12\,\ensuremath{{\rm keV_{ee}}} in COSINE-100 (see Fig.~\ref{fig:detail}B).
A cross-check incorporating the iodine QF is provided in Appendix.

\section{Data analysis}
To obtain radiation-induced scintillation events, one has to separate PMT-induced noise events that are predominantly triggered and recorded in the low-energy signal regions. 
A multivariate machine learning technique has been developed to characterize the pulse shapes to discriminate these PMT-induced noise events from radiation-induced scintillation events~\cite{COSINE-100:2020wrv, COSINE-100:2024SEL}.
To improve the discrimination power, we categorize noise types and evaluate likelihood scores in both the time domain and frequency domain, enhancing the separation between scintillation events and PMT-induced noise events.
Scintillation-rich data samples were collected by installing a $^{22}$Na source and by requiring a coincidence condition with high-energy at neighboring crystals or liquid scintillator, to tag the characteristic $\gamma$-rays of 511\,keV or 1,274\,keV (see Appendix).

Multilayer perceptron~(MLP) networks are subsequently trained with these scintillation-rich $^{22}$Na calibration data samples alongside PMT noise-dominant single-hit physics data.
Requiring less than 1\% noise contamination in the selection criteria for the MLP output score, we reached an energy threshold of 0.7\,keV~\cite{COSINE-100:2024SEL}, where a unit of keV corresponds to reconstructed energy close to the true energy, considering the nonproportionality of sodium iodide crystals~\cite{COSINE-100:2024log}.
The event selection efficiency for scintillation events is evaluated with the $^{22}$Na calibration dataset, as shown in Fig.~\ref{fig:detail}D, and is cross-checked with waveform simulation data~\cite{Choi:2024ziz} as well as with nuclear recoil calibration data~\cite{Lee:2024unz}.

Because mismodeling of the time-dependent backgrounds potentially induces modulation-like signatures~\cite{COSINE-100:2021zqh, COSINE-100:2022dvc}, we have launched further investigations into the backgrounds in the COSINE-100 detector. 
Improved energy calibration of electron recoil events through the nonproportionality measurement~\cite{COSINE-100:2024log} and improved understanding of internal contamination through internal $\alpha$ measurement~\cite{COSINE-100:2023dsf} allowed this to be achieved over extended energy ranges from 0.7\,keV to 4,000\,keV for electron/$\gamma$ background~\cite{COSINE-100:2024Background}.
The most prominent background components are $^{3}$H (half-life of 12.3\,years) and $^{210}$Pb (half-life of 22.3\,years).
Internal $^{210}$Pb is separated from surface $^{210}$Pb, contaminated on the crystal surface or the Teflon wrapping sheet, because the measured half-life of $\alpha$ particles from bulk $^{210}$Po, a decay product of $^{210}$Pb, is consistent with 22.3\,years, while the lower energy $\alpha$ particles from surface $^{210}$Po have a measured effective half-life of 33.8$\pm$8.0\,years~\cite{COSINE-100:2023dsf}.
This difference in half-lives is attributed to the continuous emanation of $^{222}$Rn by the crystal encapsulation material and contamination of the crystal surface or the Teflon wrapping sheet~\cite{COSINE-100:2023dsf}. 

\begin{figure*}[t]
    \centering
    \includegraphics[width=1.0\textwidth]{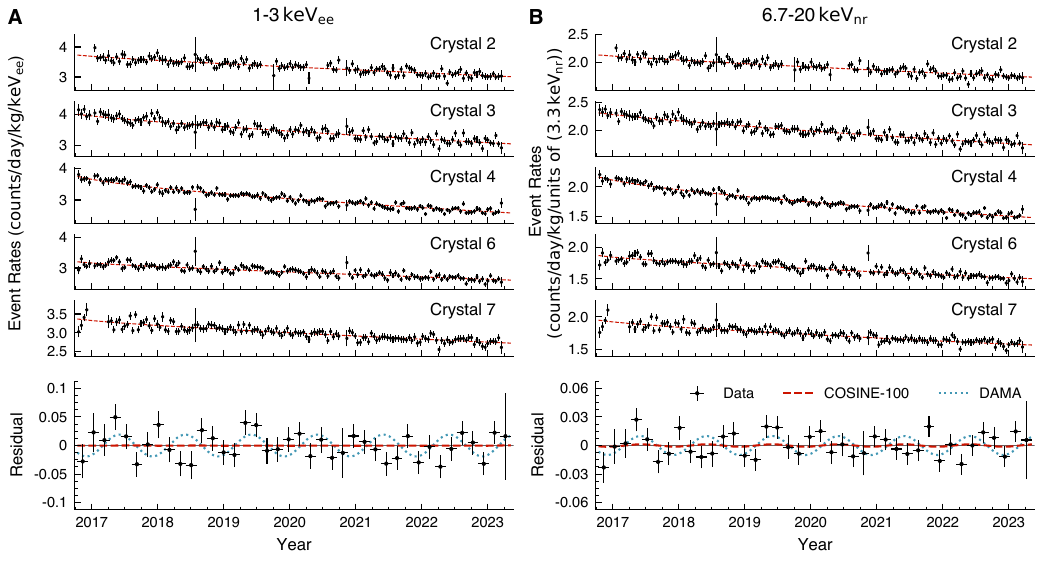}
    \caption{
        {\bf The event rates over time of each detector and modulation fit of the COSINE-100 full dataset.}
        ({\bf A}) Event rates in the 1--3\,\ensuremath{{\rm keV_{ee}}} energy region.
        ({\bf B}) Event rates in the 6.7--20\,\ensuremath{{\rm keV_{nr}}} energy region.
        The event rates (data points with 68.3\% error bars) are calculated in 15-day intervals and compared to the phase-fixed best-fit model (red dashed lines).
        In the bottom panels, the average residual event rates (data points with error bars) are shown, where the fitted background has been subtracted.
        These residuals are presented in 60-day intervals.
        Overlaid on the residuals are the fitted modulation components (red dashed lines) and the DAMA/LIBRA observations (blue dotted lines).
    }
    \label{fig:SignalFit}
\end{figure*}

To search for a modulation signal in the data, the dataset was prepared to correspond to the number of events in each 15-day time bin after applying the event selection.
The model describing the data consists of a dark matter-induced modulation signal and time-dependent backgrounds, represented by sinusoidal and exponential functions, respectively. Thus, the event rate of the $i$th sodium iodide detector is defined by:
\begin{eqnarray}
    R_{i} \left( t;\,A,\,\phi,\,\vec{C}_{i},\,\vec{\lambda}_{i} \right) = && A\,\mathrm{cos} \left( {2\pi\frac{t - \phi}{T}} \right) \nonumber \\ 
    &&+ \sum_{j}^{N_\mathrm{bkgd}} C_{ij}\exp\left({-\lambda_{ij}\,t}\right),
\label{eq:modfunc}
\end{eqnarray}
where $A$ and $\phi$ represent the amplitude and phase of the modulation signal, respectively, and $T$ is the Earth's orbital period of 365.2\,days. The initial event rate and decay constant of the $j$th background component are denoted by $C_{ij}$ and $\lambda_{ij}$, respectively, for a total of $N_{\mathrm{bkgd}}$=10 components. The evaluated detector livetime and selection efficiencies were applied to the model.

\begin{figure*}[t]
    \centering
    \includegraphics[width=1.0\textwidth]{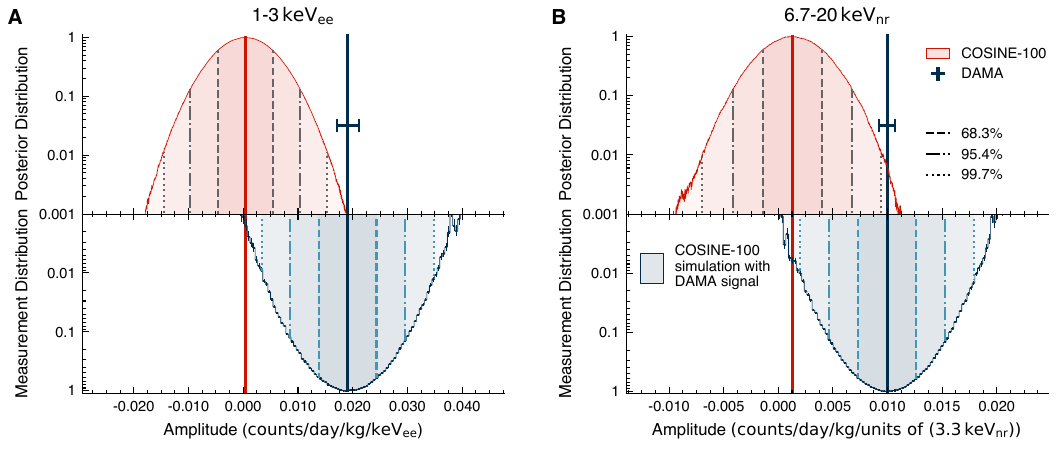}
    \caption{
        {\bf Posterior distributions of modulation amplitudes from the COSINE-100 phase-fixed fits and the expected distributions for measurements assuming the DAMA/LIBRA signals.}
        ({\bf A}) Modulation amplitude distribution in the 1--3\,\ensuremath{{\rm keV_{ee}}} region.
        ({\bf B}) Modulation amplitude distribution in the 6.7--20\,\ensuremath{{\rm keV_{nr}}} region.
        The red regions represent the posterior distributions obtained from the COSINE-100 full dataset.
        The blue regions in the lower panels show the distributions of best-fits from simulated data, assuming the expected COSINE-100 background and the observed DAMA/LIBRA signals.
        The uncertainty in DAMA/LIBRA's observations has been considered during the simulation.
        The vertical solid lines indicate the best-fit modulation amplitudes for COSINE-100 (red) and the DAMA/LIBRA best-fit values (blue) with 68.3\% errors.
        The other line styles indicate each probability region.
        The distributions are normalized to have a maximum value of unity for the comparison.
    }
    \label{fig:Posterior}
\end{figure*}

A Poisson likelihood was constructed to compare the model and data throughout every crystal simultaneously, and we utilized a Bayesian approach to determine the modulation amplitude from the likelihood.
Each background component is controlled by the corresponding $C_{ij}$, constrained by the activity and its uncertainty, including both statistical and systematic, estimated from the background modeling~\cite{COSINE-100:2024Background}.
The flat component, which is the sum of long-lived elements such as $^{40}$K, $^{238}$U, and $^{232}$Th, was free-floated to account for the non-modulating component of the dark matter-induced signals.
The decay constant $\lambda_{ij}$ for the exponential terms remains fixed during the analysis, except for that of the surface $^{210}$Pb component, which may effectively vary due to the potential emanation of $^{222}$Rn (see Appendix).

The primary analysis fixes the phase of the modulation $\phi$ to the maximum occurring on June 2, 152.5 days from the start of the calendar year, as predicted by the standard halo model~\cite{Lewin:1995rx,freese1987,Freese:2012xd,Froborg:2020tdh}, and searches for modulation signals in the energy ranges of 1--3\,\ensuremath{{\rm keV_{ee}}}, 2--6\,\ensuremath{{\rm keV_{ee}}}, and 1--6\,\ensuremath{{\rm keV_{ee}}}, where DAMA/LIBRA reported annual modulation signals~\cite{Bernabei:2021kdo}.
In addition to the same electron-equivalent energy ranges, we search for annual modulation signals in the same nuclear recoil energy region of 6.67--20\,\ensuremath{{\rm keV_{nr}}}, taking into account the different sodium nuclear recoil QFs of DAMA/LIBRA~\cite{BERNABEI1996757} and COSINE-100~\cite{Lee:2024unz} (see Appendix for the consideration of iodine QF).
This range corresponds to 2--6\,\ensuremath{{\rm keV_{ee}}} and 0.85--3.12\,\ensuremath{{\rm keV_{ee}}} for DAMA/LIBRA detectors and COSINE-100 detectors, respectively.
Although all results are summarized in Appendix, the main text focuses on the results in the energy ranges of 1--3\,\ensuremath{{\rm keV_{ee}}} and 6.67--20\,\ensuremath{{\rm keV_{nr}}}.

To prevent bias in the search for an annual modulation signal, the fitter was tested with simulated event samples prior to unblinding the full dataset.
Each simulated dataset is generated by Poisson random sampling from the modeled time-dependent background rates with assumed modulation signals.
Seventeen ensembles with modulation signals evenly varied from $-2$ to $+2$ times the modulation amplitude of DAMA/LIBRA, including the null hypothesis, were tested.
We found no bias attributed to the fitter for a wide range of modulation signals, as shown in Fig.~\ref{fig:pull_test}.
Every technique used for this analysis had been developed using 3-year or less data, without providing their time information in the region of interest~\cite{COSINE-100:2024log, COSINE-100:2024Background}.
The time-dependent information of the full dataset was unblinded after the methodology was verified based on these simulated experiments.

With a phase-fixed fit, we find best-fit modulation amplitudes of 0.0004$\pm$0.0050 and 0.0017$\pm$0.0029 counts/day/kg/\ensuremath{{\rm keV_{ee}}} in the 1--3\,\ensuremath{{\rm keV_{ee}}} and 1--6\,\ensuremath{{\rm keV_{ee}}} energy ranges, respectively.
Considering different nuclear recoil QFs, the best-fit for the 6.67--20\,\ensuremath{{\rm keV_{nr}}} range is 0.0013$\pm$0.0027\,counts/day/kg/units of (3.3\,\ensuremath{{\rm keV_{nr}}}), where 3.3\,\ensuremath{{\rm keV_{nr}}} corresponds to 1\,\ensuremath{{\rm keV_{ee}}} for the DAMA/LIBRA sodium QF of 0.3.
Figure~\ref{fig:SignalFit} shows the observed event rate over time overlaid with the phase-fixed best-fit model for 1--3\,\ensuremath{{\rm keV_{ee}}}~(A) and 6.67--20\,\ensuremath{{\rm keV_{nr}}}~(B).
For visualization purposes, we draw the normalized event rate, although the likelihood fit uses raw event counts per each 15-day time bin.
The bottom of each plot presents the residual event rates averaged over five crystals.
The fitted background components were subtracted from the event rates to calculate the residuals, where the modulation signals were not subtracted.
Red-solid lines present the best fit including the modulation signals, while the blue-dashed lines correspond to the DAMA/LIBRA's annual modulation signals~\cite{Bernabei:2021kdo}.

\begin{figure*}[t]
    \centering
    \includegraphics[width=1.0\textwidth]{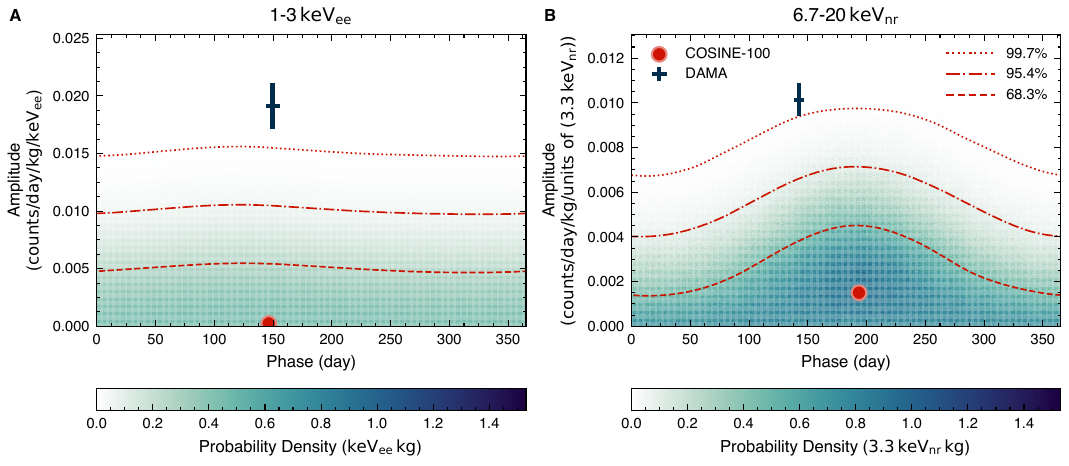}
    \caption{
        {\bf Two-dimensional posterior distributions of phase-floated modulation fits for the COSINE-100 full dataset.}
        ({\bf A}) Posterior distribution in the 1--3\,\ensuremath{{\rm keV_{ee}}} region.
        ({\bf B}) Posterior distribution in the 6.7--20\,\ensuremath{{\rm keV_{nr}}} region.
        The COSINE-100 best-fit points (red dots) and the probability contours from the posterior distributions for the phase-floated fits are compared with the best-fit amplitudes and phases reported by DAMA/LIBRA (data points with 68.3\% error bars).
    }
    \label{fig:2Dfit}
\end{figure*}

\begin{table*}[t]
    \centering
    \caption{
        {\bf Summary of phase-fixed fits in the electron recoil signal regions.}
        We summarize the modulation amplitudes obtained from the COSINE-100 full dataset in the three different electron-equivalent energy ranges of 1--3\,\ensuremath{{\rm keV_{ee}}}, 1--6\,\ensuremath{{\rm keV_{ee}}}, and 2--6\,\ensuremath{{\rm keV_{ee}}}.
        These results are compared with modulation amplitudes obtained from COSINE-100 3\,years~\cite{COSINE-100:2021zqh} and 1.7\,years~\cite{COSINE-100:2019lgn} data, ANAIS-112 3\,years data~\cite{Coarasa2024}, and DAMA/LIBRA~\cite{Bernabei:2021kdo}.
        Note that the energy ranges of the COSINE-100 full dataset and DAMA were linearly calibrated, while the previous results of COSINE-100 and ANAIS-112 adopted nonproportional calibration.
        The errors indicate the 68.3\% confidence intervals.
    }
    \label{tab:PhaseFixedFittingResults}
    \begin{tabular}{lcc}
        \\
        \hline
        Dataset                       & Energy (\ensuremath{{\rm keV_{ee}}}) & Amplitude  (counts/day/kg/\ensuremath{{\rm keV_{ee}}}) \\
        \hline
        {\bf COSINE-100 full dataset} & 1--3     & 0.0004$\pm$0.0050      \\
        DAMA/LIBRA-phase2             & 1--3     & 0.0191$\pm$0.0020      \\
        \hline
        {\bf COSINE-100 full dataset} & 1--6     & 0.0017$\pm$0.0029      \\
        COSINE-100 3 years            & 1--6     & 0.0067$\pm$0.0042      \\
        ANAIS-112 3 years             & 1--6     & -0.0013$\pm$0.0037     \\
        DAMA/LIBRA-phase2             & 1--6     & 0.0105$\pm$0.0009      \\
        \hline
        {\bf COSINE-100 full dataset} & 2--6     & 0.0053$\pm$0.0031      \\
        COSINE-100 3 years            & 2--6     & 0.0051$\pm$0.0047      \\
        COSINE-100 1.7 years          & 2--6     & 0.0083$\pm$0.0068      \\
        ANAIS-112 3 years             & 2--6     & 0.0031$\pm$0.0037      \\
        DAMA/NaI + DAMA/LIBRA         & 2--6     & 0.0100$\pm$0.0007      \\
        \hline
    \end{tabular}
\end{table*}

Figure~\ref{fig:Posterior} presents the marginalized posterior distributions of the modulation amplitudes in the 1--3\,\ensuremath{{\rm keV_{ee}}} and 6.67--20\,\ensuremath{{\rm keV_{nr}}} regions, compared with the reported modulation amplitudes from DAMA/LIBRA~\cite{Bernabei:2021kdo}.
The distributions of modulation amplitude measurements from ensembles of 300,000 simulated experiments are overlaid.
They were simulated with injected annual modulation signals, the same as DAMA/LIBRA's observation in each energy range, considering Gaussian fluctuation within the reported uncertainty.
These ensemble measurements are compared with the measured modulation amplitude from the COSINE-100 data to evaluate the hypothesis test between DAMA/LIBRA and COSINE-100.
The significance of the COSINE-100 data ruling out DAMA/LIBRA's annual modulation hypothesis is 3.57$\sigma$ for 1--3\,\ensuremath{{\rm keV_{ee}}} and 3.25$\sigma$ for 6.67--20\,\ensuremath{{\rm keV_{nr}}}, respectively.
The cross-check that accounted for the iodine QF revealed a 2.80$\sigma$ level of inconsistency with the DAMA/LIBRA claim (see Appendix).
Tables~\ref{tab:PhaseFixedFittingResults} and \ref{tab:PhaseFixedFittingResultsNr} summarize the results from the phase-fixed fits in different signal regions.
The measured modulation amplitudes from the COSINE-100 data generally agree well with null signals but disfavor the DAMA/LIBRA signal.

\begin{figure*}[t]
    \centering
    \includegraphics[width=1.0\textwidth]{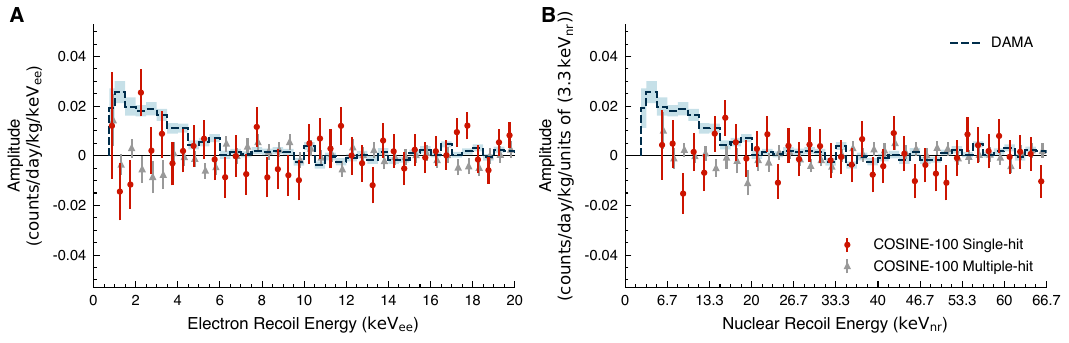}
    \caption{
        {\bf Modulation amplitude for each energy bin from the COSINE-100 full dataset.}
        ({\bf A}) Fitted modulation amplitudes in 0.75--20\,\ensuremath{{\rm keV_{ee}}}.
        ({\bf B}) Fitted modulation amplitudes in 5--66.7\,\ensuremath{{\rm keV_{nr}}}.
        The COSINE-100 single-hit (red dots) and multiple-hit (gray triangles) data are compared with the modulation amplitudes observed by the DAMA/LIBRA experiment (blue-dashed lines) with 68.3\% uncertainties (blue-shaded regions).
        The bin sizes are 0.5\,\ensuremath{{\rm keV_{ee}}}~(A) and 1.67\,\ensuremath{{\rm keV_{nr}}}~(B) except for the first bin of (A), which uses 0.25\,\ensuremath{{\rm keV_{ee}}}.
    }
    \label{fig:Amplitude}
\end{figure*}

We also search for an annual modulation signal by allowing both the amplitude and phase of the signal to vary in the fit.
The two-dimensional posterior distributions obtained from the phase-floated modulation search are shown in Fig.~\ref{fig:2Dfit}, highlighting the best-fit points of the model along with the 68.3\%, 95.4\%, and 99.7\% probability contours.
The annual modulation amplitudes and phases reported by the DAMA/LIBRA experiment are also displayed for comparison.
As seen in the figure, the results from the phase-floated modulation search agree with a null observation but disfavor the DAMA/LIBRA signal above the 3$\sigma$ level, consistent with the fixed-phase search.

\begin{table*}[t]
    \centering
    \caption{
        {\bf Phase-fixed fit result in the nuclear recoil signal region.}
        The modulation amplitudes obtained from the COSINE-100 full dataset in the nuclear recoil energy of 6.7--20\,\ensuremath{{\rm keV_{nr}}} is compared with those obtained from ANAIS-112 3\,years data~\cite{Coarasa2024} and DAMA/LIBRA~\cite{Bernabei:2021kdo}.
        The errors indicate the 68.3\% confidence intervals.
    }
    \label{tab:PhaseFixedFittingResultsNr}
    \begin{tabular}{lcc}
        \\
        \hline
        Dataset                       & Energy (\ensuremath{{\rm keV_{nr}}}) & Amplitude (counts/day/kg/units of (3.3\,\ensuremath{{\rm keV_{nr}}})) \\
        \hline
        {\bf COSINE-100 full dataset} & 6.7--20  & 0.0013$\pm$0.0027           \\
        DAMA/NaI + DAMA/LIBRA         & 6.7--20  & 0.0100$\pm$0.0007           \\
        ANAIS-112 3 years             & 6.7--20  & 0.0017$\pm$0.0025           \\
        \hline
    \end{tabular}
\end{table*}

Lastly, in Fig.~\ref{fig:Amplitude}, we present the best-fit modulation amplitude as a function of electron recoil energy at 0.75--20\,\ensuremath{{\rm keV_{ee}}} and nuclear recoil energy at 5--66.7\,\ensuremath{{\rm keV_{nr}}} for both single-hit and multi-hit events in phase-fixed fits.
We find that the $\chi^2$ test on the sideband regions of 6--20\,\ensuremath{{\rm keV_{ee}}} (20--66.7\,\ensuremath{{\rm keV_{nr}}}) single-hit and 0.75--6\,\ensuremath{{\rm keV_{ee}}} (5--20\,\ensuremath{{\rm keV_{nr}}}) multiple-hit are consistent with no modulations, with a $p$-value of 0.46 (0.82) and 0.26 (0.41), respectively.
In the signal region of single-hit 0.75--6\,\ensuremath{{\rm keV_{ee}}} (5--20\,\ensuremath{{\rm keV_{nr}}}), COSINE-100 results are consistent with no modulation, with $p$-values of 0.34 (0.26).
However, the hypothesis tests for DAMA/LIBRA's annual modulation compared to the COSINE-100 results have a $p$-value of only 0.003 (0.0003).
The COSINE-100 data strongly favor the no modulation hypothesis. 

\section{Discussion}
COSINE-100 found no evidence of annual modulation signals using the same target material as DAMA/LIBRA, with energy calibrations specifically matched to those of DAMA/LIBRA.  
This model-independent analysis of the full dataset, spanning 6.4 years of operation, significantly disfavors DAMA/LIBRA's annual modulation signals, with a 3.57$\sigma$ confidence level for electron recoil energies of 1--3\,\ensuremath{{\rm keV_{ee}}} and a 3.23$\sigma$ level for sodium nuclear recoil energies of 6.67--20\,\ensuremath{{\rm keV_{nr}}}.
With no modulation signals observed in the COSINE-100 dataset, the hypothesis that DAMA/LIBRA's annual modulation is caused by dark matter interactions is increasingly difficult to support. 
  \hfill \break

\paragraph*{\bf{Acknowledgments:}}
We thank the Korea Hydro and Nuclear Power~(KHNP) Company for providing underground laboratory space at Yangyang and the IBS Research Solution Center~(RSC) for providing high performance computing resources.
This work is supported by:
the Institute for Basic Science~(IBS) under project code IBS-R016-A1, NFEC-2019R1A6C1010027, NRF-2021R1I1A3041453, NRF-2021R1A2C3010989, NRF-2021R1A2C1013761 and RS-2024-00356960, Republic of Korea; 
NSF Grants No. PHY-1913742, United States; 
STFC Grant ST/N000277/1 and ST/K001337/1, United Kingdom; 
Grant No. 2021/06743-1, 2022/12002-7 and 2022/13293-5 FAPESP, CAPES Finance Code 001, CNPq 304658/2023-5, Brazil; 
UM grant No. 4.4.594/UN32.14.1/LT/2024, Indonesia;
S.M.L. is supported by the Hyundai Motor Chung Mong-Koo Foundation.

\paragraph*{\bf{Data and materials availability:}}
The data that support the findings of this study are available in the ``COSINE-100 Full Dataset Challenges the Annual Modulation Signal of DAMA/LIBRA'' Dryad dataset, \href{https://doi.org/10.5061/dryad.fbg79cp6t}{https://doi.org/10.5061/dryad.fbg79cp6t}.
More supplementary materials, such as visualizers, are available in the folder \texttt{COSINE100-6YearsModulation} in the public GitHub repository of the COSINE collaboration, \href{https://github.com/CUPCOSINE/PublicData}{https://github.com/CUPCOSINE/PublicData}.

\providecommand{\href}[2]{#2}\begingroup\raggedright\endgroup

\clearpage


\appendix

\renewcommand{\thefigure}{A\arabic{figure}}
\renewcommand{\thetable}{A\arabic{table}}
\renewcommand{\theequation}{A\arabic{equation}}
\renewcommand{\thepage}{A\arabic{page}}
\setcounter{figure}{0}
\setcounter{table}{0}
\setcounter{equation}{0}
\setcounter{section}{0}
\setcounter{page}{1} 

\section{Appendix}
\subsection{COSINE-100 detector and operation}

\begin{figure}[t]
    \centering
    \includegraphics[width=0.48\textwidth]{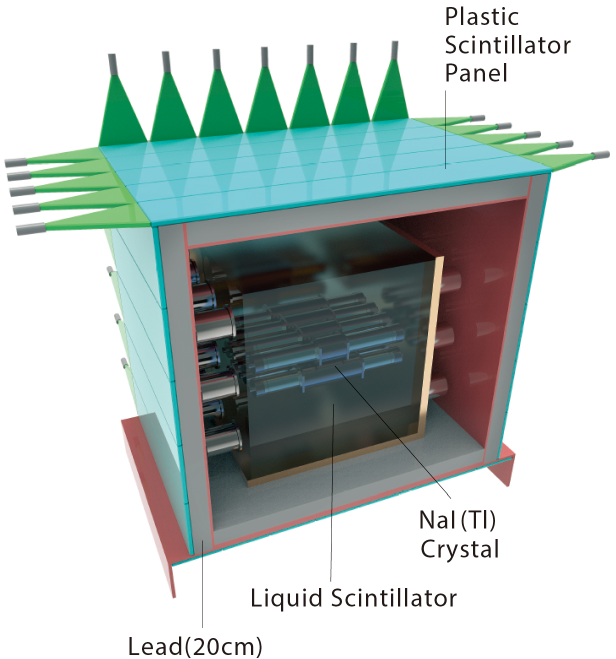}
    \caption{
        {\bf Schematic of the COSINE-100 detector.}
        The eight encapsulated sodium iodide detectors are immersed in liquid scintillator and surrounded by 20\,cm thick lead bricks and 37 plastic scintillator panels. 
    }
    \label{fig:cosinedetector}
\end{figure}

The COSINE-100 experiment was located 700\,meters below the surface at the Yangyang Underground Laboratory in eastern Korea.
A cut-out view of the detector is shown in Fig.~\ref{fig:cosinedetector}.
It began physics operation on October 21, 2016 and concluded on March 14, 2023, for relocation of the experimental site to Yemilab~\cite{Park:2024sio, Kim:2024xyd}, a new, deeper underground laboratory in Korea with approximately four times lower muon rate than Y2L. 
An upgrade of the COSINE-100 (COSINE-100U) experiment at Yemilab is currently being installed to increase light collection with an improved crystal encapsulation~\cite{Choi:2020qcj, NEON:2024Upgrade} and increased light output by operating at $-$30$^{\circ}$C~\cite{Lee:2021aoi}.

During the 6.4-year operation period, no substantial environmental abnormalities or severe instabilities in NaI(Tl) detector behavior were observed.
Physics-data-taking efficiency was achieved at a level of 95.6\%, as shown in Fig.~\ref{fig:detail}A.
A small portion of the data was affected by an exceptionally high trigger rate, typically caused by high-energy cosmic ray muons.
When such spikes were detected, the entire two-hour data file was excluded from the analysis to avoid any bias from muon-induced long-lived phosphorescence, resulting in a 1.8\% data reduction.
Additionally, a specific type of PMT-induced noise occasionally increased during certain periods, affecting only one of the eight NaI(Tl) crystal detectors at a time~\cite{COSINE-100:2021xqn, COSINE-100:2019lgn, COSINE-100:2021zqh}.
These affected periods, comprising about 2.7\% of the data, were further removed from the analysis on a detector-by-detector basis to avoid any noise-induced modulation signals.
After these filtering steps, the remaining ``high quality data'' amounted to 358.4\,kg$\cdot$years.

The small reduction in efficiency was primarily due to the occasional monthly-long calibration campaigns using $^{60}$Co and $^{22}$Na to obtain low-energy scintillation-rich samples.
Multiple-hit events recorded during these calibration campaigns with the $^{60}$Co and $^{22}$Na sources provided a large sample of Compton scattering events.
For $^{60}$Co calibration, we irradiate $\gamma$-rays using a 1\,$\mu$Ci disk source outside the liquid scintillator and required multiple-hit signals above 100\,keV from the liquid scintillator or nearby crystals.
For $^{22}$Na calibration, we prepared two stainless-steel cases suitable for the calibration tube using standard isotope solutions with approximately 50\,Bq activities.
Two calibration tubes were installed in the calibration holes in the middle of the eight crystals~\cite{Adhikari:2017esn}.
We required multiple-hit events with energy above 200\,keV considering three $\gamma$-rays of two 511\,keV and one 1,275\,keV. 

\begin{figure*}[t]
    \centering
    \includegraphics{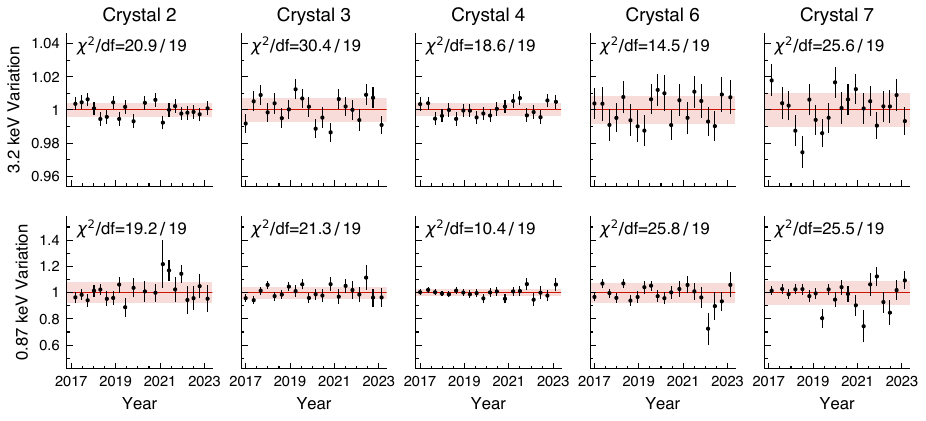}
    \caption{
        {\bf Low energy calibration stability.}
        Time variations of 0.87\,keV and 3.2\,keV peak positions across each crystal are depicted.
        Black data points with 68.3\% error bars represent measurements for different time bins, which are then divided by the overall average (red solid lines).
        The corresponding standard deviations are shown as filled areas.
        The $\chi^{2}$ values calculated between the measured points and the averaged values assess the consistency and stability of these calibrations.
    }
    \label{fig:detail_peak}
\end{figure*}

The stability of detector operation was ensured by monitoring various environmental factors~\cite{COSINE-100:2021mlj}.
In addition, we checked the stability of the low energy calibrations using two mono-energetic peaks emitted from the decay of internally contaminated $^{40}$K and $^{22}$Na.
Both isotopes decay through electron capture, emitting characteristic high-energy $\gamma$-rays and low-energy cascade x-rays of 3.2\,keV for $^{40}$K and 0.87\,keV for $^{22}$Na.
By tagging the characteristic 1,460\,keV or 1,274\,keV $\gamma$ lines from the surrounding liquid scintillator or other crystals, the two low-energy x-ray peaks can be identified.
The stability of the 0.87\,keV and 3.2\,keV lines was maintained within statistical uncertainty throughout the 6.4-year data-taking period, as seen in Fig.~\ref{fig:detail_peak}.
The increase in uncertainties of the 0.87\,keV peak from $^{22}$Na is expected due to its 2.6-year half-life, which causes a decrease in statistics over time.
In contrast, no such effect is observed in the 3.2\,keV peak from $^{40}$K, which has a much longer half-life of $10^9$\,years.

\subsection{Energy calibration for comparison with DAMA/LIBRA}
In the low-energy electron recoil calibration in the range of 0--100\,keV, the DAMA/LIBRA experiment studied various external $\gamma$ sources and internal x-rays or $\gamma$ lines~\cite{Bernabei:2008yh}.
External $\gamma$ sources are $^{241}$Am and $^{133}$Ba, which provided 30.4\,keV, 59.5\,keV, and 81.0\,keV peaks.
Internal x-rays from $^{40}$K provided 3.2\,keV peak and internal $^{125}$I and $^{129}$I provided 39.6\,keV, 40.4\,keV, and 67.3\,keV lines with $\beta$ or Auger electrons.
The linear fit to the calibration points was adopted as shown in Fig.~20 of Ref.~\cite{Bernabei:2008yh}.
From this figure, one can see that their linear fit started from the origin and passed through the center of the 59.5\,keV calibration point, possibly due to large calibration samples of $^{241}$Am regularly used for running conditions~\cite{Bernabei:2008yh}.
Although we found nonproportionality of scintillation light output from our sodium iodide crystals as in Fig.~\ref{fig:detail}C~\cite{COSINE-100:2024log}, we employed a calibration method similar to DAMA/LIBRA for this analysis, with a linear fit starting from the origin and using the 59.5\,keV peak, and defining \ensuremath{{\rm keV_{ee}}} as the unit keV electron-equivalent energy from this linear calibration.

To test DAMA/LIBRA's signal under the assumption of it being a nuclear recoil signal, we accounted for the QF difference reported by the DAMA experiment~\cite{BERNABEI1996757} and the COSINE-100 experiment~\cite{Lee:2024unz}.
As the nuclear recoil signal of iodine atoms is strongly constrained by other dark matter search experiments, such as LZ~\cite{LZ:2022ufs}, XENON~\cite{XENONCollaboration:2023orw}, PandaX-4T~\cite{PandaX-4T:2021bab}, and KIMS~\cite{Kim:2011je}, we focused on sodium nuclei only.
The QF of a thallium-doped sodium iodide crystal from the same ingot of COSINE-100 crystal-6 and crystal-7 was measured independently using a mono-energetic neutron generator through the deuteron-deuteron nuclear fusion reaction.
As shown in Fig.~\ref{fig:detail}B, the sodium QF from the COSINE-100 measurement~\cite{Lee:2024unz} was approximately half of the DAMA measurement~\cite{BERNABEI1996757} in the region of interest.

\subsection{Event selection}
As DAMA/LIBRA reported the results with a 0.75\,\ensuremath{{\rm keV_{ee}}} energy threshold~\cite{Bernabei:2021kdo}, we have reduced the analysis threshold from 1\,keV~\cite{COSINE-100:2020wrv} to 0.7\,keV by applying multivariate machine learning techniques~\cite{COSINE-100:2024SEL}.
The 0.7\,keV threshold considers nonproportionality of sodium iodide crystals~\cite{Lee:2024unz} and corresponds to 0.54\,\ensuremath{{\rm keV_{ee}}} with DAMA-like linear calibration.
There are a couple of updates from previous selection~\cite{COSINE-100:2020wrv}, as we used a month-long $^{22}$Na calibration data instead of $^{60}$Co calibration data with higher probabilities of the scintillation-rich events from three coincident $\gamma$-radiations.
Various likelihood and mean time-related parameters were developed for the separation of specific types of noise events.
Instead of using a boosted decision tree, we employed multilayer perceptrons~(MLPs), with two hidden layers, utilizing the \textsc{TMVA} package~\cite{TMVA:2007ngy} of CERN \textsc{ROOT} framework~\cite{Brun:1997pa}.
The updated event selection is described in detail elsewhere~\cite{COSINE-100:2024SEL}.

We have evaluated the selection efficiency using $^{22}$Na calibration data as a function of energy in the \ensuremath{{\rm keV_{ee}}} unit and presented in Fig.\ref{fig:detail}D.
Simulated waveforms~\cite{Choi:2024ziz} verified the efficiency from $^{22}$Na calibration data.
Small deviations in the selection efficiency impacted the background modeling process, which in turn affected the final activity estimations.
We quantified this impact and included it as a systematic uncertainty in the activity values.
The responses of the MLPs were compared across distinct data-taking periods, showing a consistent distribution of scores.
The updated event selection process and its validation are described in detail elsewhere~\cite{COSINE-100:2024SEL}.

\begin{table*}[t]
    \centering
    \caption{
        {\bf The expected \revised{and fitted} event rates for various time-dependent background components.}
        The expected event rates in the 1--3\,\ensuremath{{\rm keV_{ee}}} region as of October 21, 2016, calculated from background modeling of the COSINE-100 data~\cite{COSINE-100:2024Background}, are listed for different time-dependent components, \revised{together with the phase-fixed fitting results}.
        The errors represent the 68.3\% confidence intervals.
        \revised{Note that the flat component was not constrained during the fitting.}
    }
    \label{tab:Components}
    \begin{tabular}{rccl}
        \\
        \hline
        & Expected Initial Rate & Fitted Initial Rate & Half-life \\
        Component & (counts/day/kg/\ensuremath{{\rm keV_{ee}}}) & (counts/day/kg/\ensuremath{{\rm keV_{ee}}}) & (years) \\
        \hline
        Total & $3.67\pm0.34$ & $3.59 \pm 0.15$ & \\
        $^{3}$H & $1.31\pm0.32$ & $1.33 \pm 0.08$ & 12.3 \\
        Surface $^{210}$Pb & $1.13\pm0.10$ & $1.05\pm0.08$ & 33.8$\pm$8.0 \\
        Internal $^{210}$Pb & $(9.46\pm0.79) \times 10^{-1}$ & $(9.13\pm0.72) \times 10^{-1}$ & 22.3 \\
        Flat & $(1.82\pm0.13) \times 10^{-1}$ & $(1.83\pm0.56) \times 10^{-1}$ & \\
        $^{109}$Cd & $(4.48\pm0.65) \times 10^{-2}$ & $(5.22\pm0.60) \times 10^{-2}$ & 1.26 \\
        $^{\rm 127m}$Te & $(2.65\pm0.61) \times 10^{-2}$ & $(2.82\pm0.50) \times 10^{-2}$ & 0.29 \\
        $^{113}$Sn & $(2.04\pm0.39) \times 10^{-2}$ & $(2.35\pm0.36) \times 10^{-2}$ & 0.31 \\
        $^{22}$Na & $(5.68\pm1.61) \times 10^{-3}$ & $(6.28\pm1.46) \times 10^{-3}$ & 2.60 \\
        $^{\rm 121m}$Te & $(3.15\pm0.65) \times 10^{-3}$ & $(3.24\pm0.61) \times 10^{-3}$ & 0.45 \\
        $^{60}$Co & $(4.38\pm0.28) \times 10^{-5}$ & $(4.38\pm0.28) \times 10^{-5}$ & 5.27 \\
        \hline
    \end{tabular}
\end{table*}

\subsection{Simulated experiments}
The simulated datasets were prepared using a time-dependent background model of each crystal for a 15-day time bin.
The initial event rate for each background component was randomly drawn from a Gaussian distribution with mean and standard deviation being the activity and error estimated from the background modeling~\cite{COSINE-100:2024Background}.
In addition, the decay constant of surface $^{210}$Pb had a Gaussian random fluctuation from the measured value of 33.8$\pm$8.0\,years~\cite{COSINE-100:2023dsf}.
Several values of $A$, the modulation amplitude, are chosen for the simulated experiments.
The expected number of events in the $k$th time bin of the $i$th detector, $\hat{E}_{ik}$, is calculated as
\begin{equation}
    \hat{E}_{ik} = R_{i} \left(t_{k}\right) \, \Delta t \, m_{i} \, \Delta E \, \varepsilon_{ik}^{\mathrm{(livetime)}} \, \varepsilon_{i}^{\mathrm{(selection)}},
\end{equation}
where $R_{i}$ is the event rate model in Eq.~\ref{eq:modfunc}, $m_{i}$ is the mass of $i$th detector, and $\Delta t$ and $\Delta E$ are the widths of the time bin and energy range, respectively.
The efficiency of livetime and event selection were also accounted for.
The dataset is generated by randomizing the number of events in each time bin via a Poisson distribution.

The model is compared to each dataset via a Poisson binned likelihood function,
\begin{equation}
    \mathcal{L} = \prod_{i}\prod_{k} \frac{\hat{E}_{ik}^{n_{ik}}\,e^{-\hat{E}_{ik}}}{n_{ik}!}\,\pi_{i},
\end{equation}
where $n_{ik}$ is the number of measured or generated events in $k$th time bin of $i$th detector.
We used Gaussian constraints for the nuisance parameters,
\begin{eqnarray}    
    \pi_{i}\left(\vec{C}_{i},\,\vec{\lambda}_{i}\right) = && \prod_{j}\exp{\left[-\frac{1}{2}\left(\frac{C_{ij}-\mu_{ij}^{C}}{\sigma_{ij}^{C}}\right)^2\right]} \nonumber \\
    && \times
    \exp{\left[-\frac{1}{2}\left(\frac{\lambda_{ij}-\mu_{ij}^{\lambda}}{\sigma_{ij}^{\lambda}}\right)^2\right]}
\end{eqnarray}
where $C_{ij}$ and $\lambda_{ij}$ are the initial event rate and the decay constant, respectively, of the $i$th crystal $j$th background.
Ten time-dependent background components are considered: internal $^{210}$Pb, surface $^{210}$Pb, $^{3}$H, $^{\rm 127m}$Te, $^{\rm 121m}$Te, $^{113}$Sn, $^{109}$Cd, $^{22}$Na, $^{60}$Co, and a long-lived flat component~\cite{COSINE-100:2021zqh}.
For the mean $\mu_{ij}^{C}$ and standard deviation $\sigma_{ij}^{C}$ of the Gaussian constraints, we use the values and uncertainties estimated from background modeling~\cite{COSINE-100:2024Background}.
The values in the 1--3\,\ensuremath{{\rm keV_{ee}}} range are provided in table~\ref{tab:Components} as an example.
With the exception of surface $^{210}$Pb, known decay constants are used as $\mu_{ij}^{\lambda}$, and $\lambda_{ij}$ are fixed to $\mu_{ij}^{\lambda}$ while the surface $^{210}$Pb has $\mu_{ij}^{\lambda}$=33.8\,years and $\sigma_{ij}^{\lambda}$=8.0\,years from the $\alpha$ background study~\cite{COSINE-100:2023dsf}.
Those nuisance parameters are marginalized out to obtain the posterior probability density function~(PDF) for the modulation amplitude:
\begin{equation}
    \mathcal{P}(A,\,\phi) = N\int\mathcal{L}(A,\,\phi;\,\mathbf{C},\,\mathbf{\lambda})\,d\mathbf{C}\,d\mathbf{\lambda},
\label{eq:posterior}
\end{equation}
where $N$ is a normalization factor.
The prior probability for the modulation amplitude is incorporated into $N$ because we choose a flat prior.
An analysis tool was developed to obtain the posterior PDF by using the Metropolis-Hastings~\cite{ref:metropoils, ref:hastings} Markov chain Monte Carlo~(MCMC) algorithm~\cite{ref:mcmc1, ref:mcmc2} and this tool has already been used for various searches of dark matter in the COSINE-100 experiment~\cite{COSINE-100:2021xqn, COSINE-100:2021poy, COSINE-100:2023tcq}.

\begin{figure*}[t]
    \centering
    \includegraphics[width=1.0\textwidth]{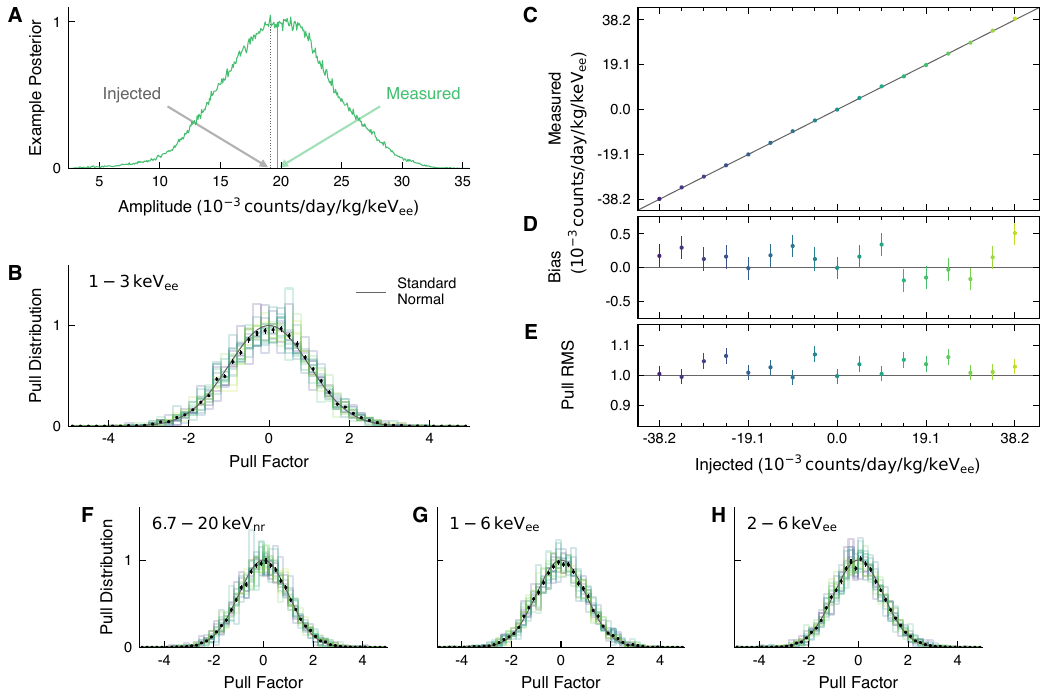}
    \caption{
        {\bf Pull test in 1--3\,\ensuremath{{\rm keV_{ee}}}.}
        ({\bf A}) An example of the posterior distribution from a simulated experiment, where the DAMA/LIBRA signal was assumed.
        ({\bf B}) The distribution of pull factors. Each color represents an injected modulation amplitude, and black dots represent their accumulated distribution, which is consistent with the standard normal distribution (grey solid curve).
        ({\bf C}) The measured modulation amplitudes as a function of the injected amplitudes.
        ({\bf D}, {\bf E}) Bias and root-mean-square of pull factors that follow the standard normal distributions independently of the injected modulation amplitudes within the 68.3\% error ranges.
        ({\bf F-H}) The distributions of pull factors for different regions of interest, which are also consistent with the standard normal distribution.
    }
    \label{fig:pull_test}
\end{figure*}

Figure~\ref{fig:pull_test}A shows a posterior PDF obtained by Eq.~\ref{eq:posterior} as an example of a simulated experiment.
As shown in the figure, the median and 1$\sigma$ confidence interval can be estimated, indicating the modulation amplitude and its uncertainty.
For the bias test, a full factor $z$ can be defined as,
\begin{equation}
    z = \frac{m_{A} - I_{A}}{\sigma_{A}},
\end{equation}
where $m_{A}$ and $\sigma_{A}$ are the measured modulation amplitude and its uncertainty from the simulated experiment, while $I_{A}$ is the input modulation amplitude of the simulated experiment.
The full factor should be identical to the standard normal distribution if there is no bias, and Fig.~\ref{fig:pull_test}B shows the distribution from various simulated experiments in 1--3\,\ensuremath{{\rm keV_{ee}}} (F-H for other energy regions).
The results of the bias test with varying modulation amplitude can be seen in Fig.~\ref{fig:pull_test}C--E, where the observed bias is negligible, less than 0.5\%, compared to the modulation amplitude reported by the DAMA/LIBRA experiment.

\subsection{Data fit and significance}
\begin{figure*}[t]
    \centering
    \includegraphics[width=1.0\textwidth]{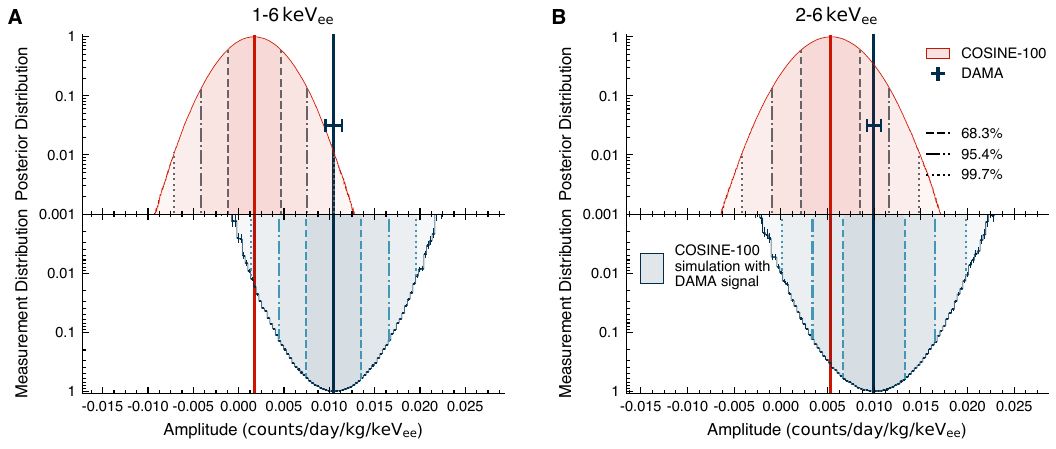}
    \caption{
        {\bf Posterior distributions of modulation amplitudes from the COSINE-100 phase-fixed fits and the expected distributions for measurements assuming the DAMA/LIBRA signals.}
        ({\bf A}) Modulation amplitude distribution in the 1--6\,\ensuremath{{\rm keV_{ee}}} region.
        ({\bf B}) Modulation amplitude distribution in the 2--6\,\ensuremath{{\rm keV_{ee}}} region.
        The red regions represent the posterior distributions obtained from the COSINE-100 full dataset.
        The blue regions in the lower panels show the distributions of best-fits from simulated data, assuming the expected COSINE-100 background and the observed DAMA/LIBRA signals.
        The uncertainty in DAMA/LIBRA's observations has been considered during the simulation.
        The vertical solid lines indicate the best-fit modulation amplitudes for COSINE-100 (red) and the DAMA/LIBRA best-fit values (blue) with 68.3\% errors.
        The other line styles indicate each probability region.
        The distributions are normalized to have a maximum value of unity for the comparison.
    }
    \label{fig:Posterior_supplement}
\end{figure*}

We searched for the electron recoil signal in the following three energy ranges: 1--3\,\ensuremath{{\rm keV_{ee}}}, 1--6\,\ensuremath{{\rm keV_{ee}}}, and 2--6\,\ensuremath{{\rm keV_{ee}}}, and the nuclear recoil signal in 6.67--20\,\ensuremath{{\rm keV_{nr}}}, which corresponds to 2--6\,\ensuremath{{\rm keV_{ee}}} in the DAMA/LIBRA detector.
The same Bayesian approach discussed and tested with the simulated dataset was applied, where $n_{ik}$, the number of events in the time bin, was obtained from the COSINE-100 full dataset.
A 15-day time bin was adopted, and no systematic effect was observed by changing the time bin width from 1 day to 60 days.
Ten MCMC chains were fitted with 100,000,000 samples generated for each chain to make the result robust to the initial condition.
All the chains converged successfully with consistent measurements within the sensitivity of the experiment.
The posterior distributions of the modulation amplitude in phase-fixed fits are presented after accumulating along the 10 chains, in Fig.~\ref{fig:Posterior} (1--3\,\ensuremath{{\rm keV_{ee}}} and 6.67--20\,\ensuremath{{\rm keV_{nr}}}) and Fig.~\ref{fig:Posterior_supplement} (1--6\,\ensuremath{{\rm keV_{ee}}} and 2--6\,\ensuremath{{\rm keV_{ee}}}), showing clear Gaussian shapes.

To test the claim of the dark matter-induced annual modulation signals observed by DAMA/LIBRA, we performed simulated experiments assuming COSINE-100 backgrounds with modulation amplitudes the same as the DAMA/LIBRA experiment for 300,000 simulated datasets for each energy range.
For each simulated experiment, the injected modulation amplitude $I_{A}$ was randomly selected from a Gaussian distribution with mean and variation from the modulation amplitude and the associated uncertainty of the DAMA/LIBRA experiment~\cite{Bernabei:2021kdo}.
Each simulated experiment generated 1,000,000 MCMC samples and found the best-fit result.
Distributions of the best-fit modulation amplitude assuming COSINE-100 backgrounds and DAMA/LIBRA's signals are shown in Fig.~\ref{fig:Posterior} (1--3\,\ensuremath{{\rm keV_{ee}}} and 6.67--20\,\ensuremath{{\rm keV_{nr}}}) and Fig.~\ref{fig:Posterior_supplement} (1--6\,\ensuremath{{\rm keV_{ee}}} and 2--6\,\ensuremath{{\rm keV_{ee}}}).
Distributions from 300,000 simulated experiments also followed the Gaussian distribution. Our measurements (red vertical lines) were significantly away from the distributions of the simulated experiments assuming DAMA/LIBRA's annual modulation signals, especially for 1--3\,\ensuremath{{\rm keV_{ee}}} and 6.67--20\,\ensuremath{{\rm keV_{nr}}}, with above 3$\sigma$ deviations.
The measured modulation amplitudes from the COSINE-100 full dataset are summarized in Tables~\ref{tab:PhaseFixedFittingResults} and \ref{tab:PhaseFixedFittingResultsNr}, along with the modulation amplitudes measured by COSINE-100 3 years data~\cite{COSINE-100:2021zqh}, DAMA/LIBRA~\cite{Bernabei:2021kdo}, and ANAIS-112~\cite{Coarasa2024}.

\begin{figure*}[t]
    \centering
    \includegraphics[width=1.0\textwidth]{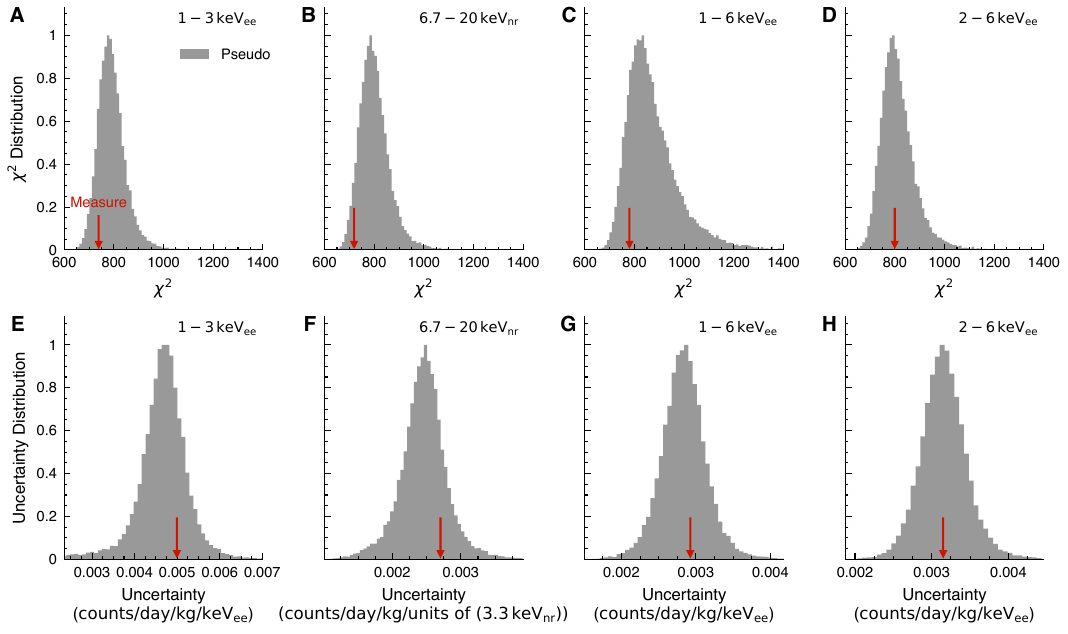}
    \caption{
        {\bf $\chi^2$ goodness-of-fits and uncertainties of the COSINE-100 data compared to the simulated experiments.}
        ({\bf A--D}) $\chi^2$s and ({\bf E--H}) the uncertainties measured from the COSINE-100 full dataset (red arrows) are compared with distributions expected from the simulated experiments with no annual modulation signals in the considered energy ranges.
        In all cases, results from the COSINE-100 data are well within 2$\sigma$ of the distributions from the 25,000 simulated experiments.
    }
    \label{fig:pseudo_uncertainty}
\end{figure*}

To check the compatibility of the model with the data, we calculated the $\chi^{2}$ of the number of events from the best-fit expectation.
The distributions of the $\chi^{2}$ were obtained by 25,000 simulated experiments assuming no modulation signal and are shown in Fig.~\ref{fig:pseudo_uncertainty}A--D.
The $\chi^{2}$ values calculated from the COSINE-100 full data fit are also displayed as red arrows, which fell within the expected distribution.
A similar check for the uncertainty of the modulation amplitude was also performed and found that uncertainties from the COSINE-100 full data fit are in agreement with uncertainty distributions obtained from the 25,000 simulated experiments as shown in Fig.~\ref{fig:pseudo_uncertainty}E--H.

\begin{figure*}[t]
    \centering
    \includegraphics[width=1.0\textwidth]{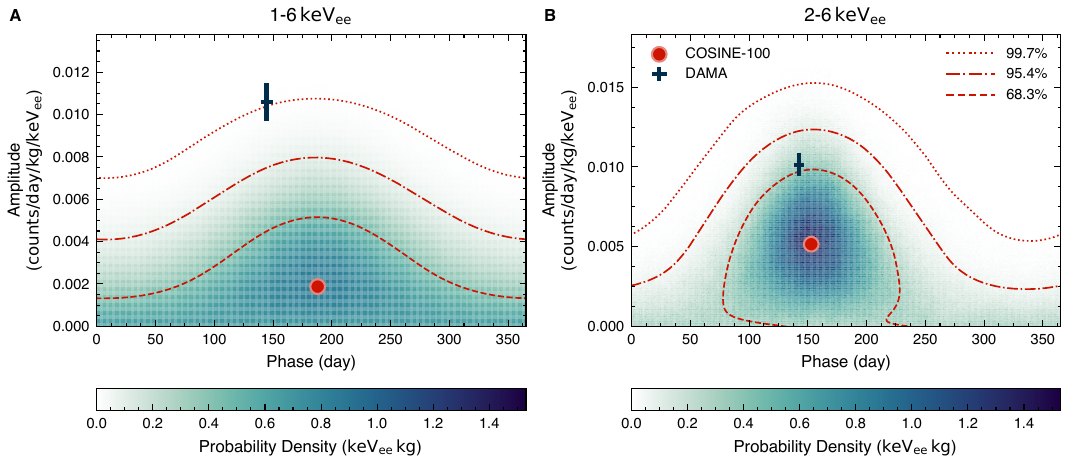}
    \caption{
        {\bf Two-dimensional posterior distributions of phase-floated modulation fits for the COSINE-100 full dataset.}
        ({\bf A}) Posterior distribution in the 1--6\,\ensuremath{{\rm keV_{ee}}} region.
        ({\bf B}) Posterior distribution in the 2--6\,\ensuremath{{\rm keV_{ee}}} region.
        The COSINE-100 best-fit points (red dots) and the probability contours from the posterior distributions for the phase-floated fits are compared with the best-fit amplitudes and phases reported by DAMA/LIBRA (data points with 68.3\% error bars).
    }
    \label{fig:2dfit_supplement}
\end{figure*}

Annual modulation signals with arbitrary phases were also searched for.
In these fits, the phase term $\phi$ was set to follow a flat prior in the same way as the modulation amplitude term $A$ in Eq.~\ref{eq:posterior}.
The same number of chains and MCMC samples were fitted as in the phase-fixed fits, and are presented in Fig.~\ref{fig:2Dfit} (1--3\,\ensuremath{{\rm keV_{ee}}} and 6.67--20\,\ensuremath{{\rm keV_{nr}}}) and Fig.~\ref{fig:2dfit_supplement} (1--6\,\ensuremath{{\rm keV_{ee}}} and 2--6\,\ensuremath{{\rm keV_{ee}}}).
The probability density regions with 68.3\%, 95.5\%, and 99.7\% were estimated after smoothing the posterior distributions using Gaussian kernel density estimation~\cite{Scott1992,Scottbook}.
A similar significance of disfavoring the DAMA/LIBRA modulation signals without 2--6\,\ensuremath{{\rm keV_{ee}}}, as in the fixed-phase search, is obtained as shown in the two figures.

\subsection{Consideration of iodine recoil}
To address the QF inconsistency problem, the annual modulation in the 6.67--20\,\ensuremath{{\rm keV_{nr}}} range has been reported throughout this article, where \ensuremath{{\rm keV_{nr}}} refers to nuclear recoil energy using the sodium QF values measured by DAMA/LIBRA and COSINE-100, respectively.
A similar comparison can be made using the iodine QF.
Although xenon-based experiments have strongly constrained the heavy-nucleus recoil interpretation of the DAMA/LIBRA claim, it is, in principle, still possible to hypothesize a particle with different couplings to iodine and xenon.
To test this possibility, we report the modulation amplitude after calibrating the energy scale using the iodine QF values measured by each experiment in this section.

DAMA/LIBRA reported a constant iodine QF value of 0.09~\cite{BERNABEI1996757}, which implies that their 2--6\,\ensuremath{{\rm keV_{ee}}} energy range corresponds to 22.2--66.7\,\ensuremath{{\rm keV_{nr,~I}}}, where \ensuremath{{\rm keV_{nr,~I}}} denotes the iodine recoil-equivalent energy.
This range is then converted into the visible energy range of 1.23--4.31\,\ensuremath{{\rm keV_{ee}}} for COSINE-100, using iodine QF values of 5--6\% measured in the same recoil energy range~\cite{Lee:2024unz}.
The same statistical method was applied to extract the annual modulation signal from the data.
The average event selection efficiency was found to be greater than 98\% in this energy range.

The annual modulation signal in the 22.2--66.7\,\ensuremath{{\rm keV_{nr,~I}}} range was measured to be 0.0015$\pm$0.0029 counts/day/kg/units of (11\,\ensuremath{{\rm keV_{nr,~I}}}).
Here, 11\,\ensuremath{{\rm keV_{nr,~I}}} corresponds to 1\,\ensuremath{{\rm keV_{ee}}} based on the iodine QF used by DAMA/LIBRA.
This result is consistent with the null signal hypothesis within 1$\sigma$ confidence level, while showing a large discrepancy from the DAMA/LIBRA result of 0.0100$\pm$0.0007 counts/day/kg/units of (11\,\ensuremath{{\rm keV_{nr,~I}}}).

\begin{figure*}[t]
    \centering
    \includegraphics{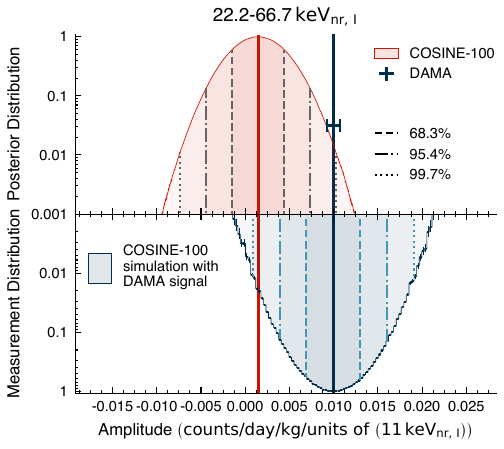}
    \caption{
        {\bf Modulation amplitude distribution in the 22.2--66.7\,\ensuremath{{\rm keV_{nr, ~ I}}} region, considering the iodine QF difference.}
        The red regions represent the posterior distributions obtained from the COSINE-100 full dataset.
        The blue regions in the lower panels show the distributions of best-fits from simulated data, assuming the expected COSINE-100 background and the observed DAMA/LIBRA signals.
        The uncertainty in DAMA/LIBRA's observations has been considered during the simulation.
        The vertical solid lines indicate the best-fit modulation amplitudes for COSINE-100 (red) and the DAMA/LIBRA best-fit values (blue) with 68.3\% errors.
        The other line styles indicate each probability region.
        The distributions are normalized to have a maximum value of unity for the comparison.
    }
    \label{fig:PhaseFixedFittingPlot_Iodine}
\end{figure*}

The distribution of the best-fit modulation amplitudes was obtained from 300,000 simulated datasets, assuming the COSINE-100 background and the DAMA/LIBRA signal, including its uncertainty.
As shown in Fig.~\ref{fig:PhaseFixedFittingPlot_Iodine}, 2.80$\sigma$ significance was observed from this best-fit distribution in comparison with the DAMA/LIBRA result.
This cross-check poses another challenge to the iodine recoil interpretation of the DAMA/LIBRA signal.

\end{document}